\documentstyle[epsfig, amsfonts]{article}
\begin{document}

\title{ Localized Solutions of the Non-Linear
 Klein-Gordon Equation in Many Dimensions}

\author{M.V.~Perel, I.V.~Fialkovsky
\footnote{Physics Faculty, St. Petersburg University, Ulyanovskaya
1-1, Petrodvorets, St. Petersburg, 198904, Russia; E-mail:
ifialk@gmail.com, perel@mph.phys.spbu.ru}}

\maketitle
\begin{abstract}We present a new complex non-stationary particle-like
solution of the non-linear Klein-Gordon equation with several
spatial variables. The construction is based on reduction to an
ordinary differential equation.\end{abstract}

The problem of finding or proving the existence of  localized
solutions of the non-linear Klein-Gordon equation in many spatial
dimensions was discussed in many papers from mathematical,
physical and numerical points of view \cite{Strauss}-\cite{Xin}.
The book \cite{Maslov} is devoted to complex asymptotic solutions
of non-linear equations. We use  the approach to construction of
localized solutions of linear equations
\cite{Kiselev-Perel,Perel-Fialkov}.

Here we give a method of calculating  complex localized solutions
of the non-linear Klein-Gordon equation. For moderate time this
solution has simple explicit exponentially decreasing asymptotic
behavior outside some area moving with the group speed. The first
term of this asymptotics is the exact solution of the linear
Klein-Gordon equation  presented earlier in \cite{Perel-Fialkov}
which decrease exponentially away from the point moving along the
straight line. Inside the moving area this solution can be found
numerically from an ordinary differential equation of some complex
variable depending on the time and spatial coordinates.

\paragraph{ Particle-like solution on the linear Klein-Gordon equation in two dimensions.}
We consider the linear Klein-Gordon equation with constant
coefficients
\begin{equation}
c^{-2} v_{tt} -  \triangle v + m^2 v = 0, \quad \triangle v =
v_{xx} + v_{zz}. \label{K-H}
\end{equation}
The equation (\ref{K-H}) has the solution depending on a single
variable $s$ (see \cite{Perel-Fialkov})
\begin{equation}
v = {\exp{(i m s)} \over s } \label{U}
\end{equation}
with $s$ depending on the spatial coordinates and time as follows
\begin{equation}
s  =   i \sqrt{ (z - i k b)^2 + x^2 - (c t - i {\omega \over c}
b)^2 }
   =   i \sqrt{m^2 b^2  + x^2 + z^2 - c^2 t^2 +
     2 i b( \omega t - k z) }.
\label{S}
\end{equation}
Here $k, b$ are free parameters and  $\omega = c \sqrt{k^2 +
m^2}.$

It is shown in \cite{Perel-Fialkov} that the solution (\ref{U})
has finite energy when $b$ and $k$ are real and ${\rm Im} s
> 0$. If the time is small enough $|t| \ll {b m^2/\omega}$ than the
solution decreases exponentially for $|x| \to \infty$ and $|z| \to
\infty$. If $|x| \ll  b m $ and $|z| \ll \min{(b
m^2/k, b m )}$ then the expansion of the form
\begin{equation}
i m s  \sim  - b m^2 - {( z - v_{gr} t)^2 \over
\Delta_{\parallel}^2 }
    - { x^2 \over {\Delta}_{\perp}^2 } - i (\omega t - k z)
\label{asym-S}
\end{equation}
holds.  We use the following notations $ \Delta_{\parallel} =
\sqrt{2b} m c /\omega$,  $\Delta_{\perp} = \sqrt{2 b}$. From
(\ref{asym-S}) and (\ref{U}) it follows that
 the solution represents a wave packet with the Gaussian
envelope filled with oscillations. It moves with the group speed
$v_{gr} = {d \omega / d k}$ in the positive direction of the $z$
axis. This is demonstrated by the numerical calculations of the
solution (\ref{U}) in successive times, see Fig.1 where the
results are presented for the parameters $m=5$, $c=1$, $k=2$,
$b=15$ in the conventional units.


\begin{figure}
\begin{center}
\epsfig{file=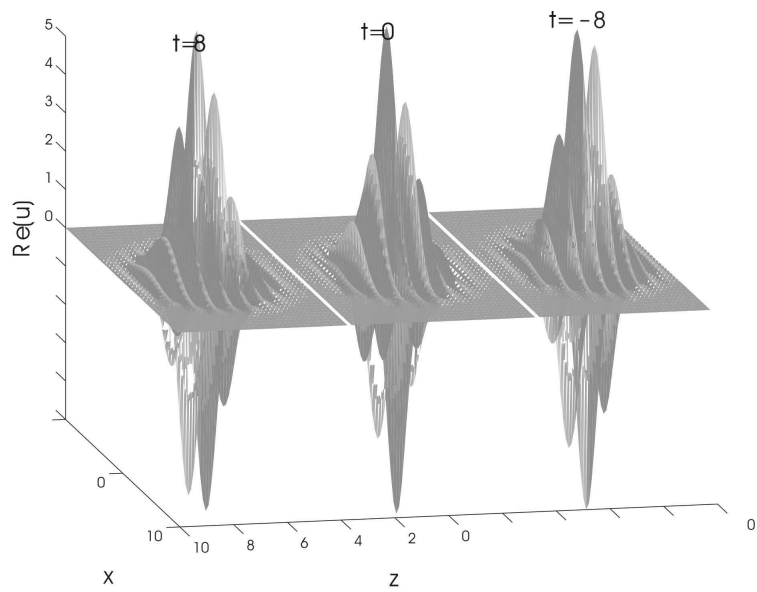,width=10truecm}
\end{center}
\caption{Particle-like solution on the
Klein-Gordon equation in the successive times in conventional
units}
\end{figure}


\paragraph{ Non-linear Klein-Gordon equation in two spacial dimensions.}

We search now the solution on the non-linear
Klein-Gordon equation in two dimensional space
\begin{equation}
c^{-2} u_{tt} -  \triangle u + f(u) = 0
 \label{equa}
\end{equation}
depending on the spatial coordinates and time only through
 the complex variable $s$ defined by (\ref{S}).
Then the partial differential equation (\ref{equa}) reduces to the ordinary
differential equation
\begin{equation}
u_{ss} + {2 \over s} u_s + f(u) = 0.
\label{ord-n-KH}
\end{equation}
Choosing for the sake of definiteness the function $f(u)$ as
follows
\begin{equation}
f(u)= m^2 u + \gamma u^{3}, \quad \gamma = const \label{f},
\end{equation}
we prove  that there exists the exact solution on non-linear
equation (\ref{equa}) having an estimate
\begin{equation}
u(s) = C {\exp{(i m s)} \over s} ( 1 + O (q \exp{(-2a)}  ),\quad C=const,
 \label{estimate}
\end{equation}
if $q \exp{(-2a)}$ is small enough, where
\begin{equation}
 {\rm Re} (ims) \le (-a)< 0 , \quad q = \gamma C^2/(m |S|).
\label{cond}
\end{equation}
  We use here the technique of
integral equations. In conditions of the validity of
(\ref{asym-S}) the inequality (\ref{cond}) can be written as
follows
\begin{equation}
{( z - v_{gr} t)^2 \over \Delta_{\parallel}^2 }
    + { x^2 \over {\Delta}_{\perp}^2 } \ge a - bm^2.
\label{vicinity}
\end{equation}
The asymptotics (\ref{estimate}) is valid for the solution of
(\ref{ord-n-KH}) outside the ellipse (\ref{vicinity}). Inside the
ellipse (\ref{vicinity}) the equation (\ref{ord-n-KH}) should be
solved numerically.

\paragraph{Non-linear Klein-Gordon equation in many dimensional space.}

The Klein-Gordon equation in many dimensional space
\begin{equation}
c^{-2}u_{tt} - \triangle u + f(u) =0, \quad \triangle u = u_{x_1x_1} +
u_{x_2 x_2} + \ldots + u_{x_nx_n},
\label{u-many}
\end{equation}
can be treated analogously to the case of two dimensional space.
Seeking the solution of (\ref{u-many}) $u$  as the function of the
single complex variable $s$
\begin{equation}
s  =   i \sqrt{ (x_1 - i k b)^2 + x_2^2 + \ldots + x_n^2 - (c t -
i {\omega \over c} b)^2 }. \label{S-many}
\end{equation}
we obtain the ordinary differential equation
\begin{equation}
u_{ss} + {n \over s} u_s + f(u) = 0.
\label{ord-n-KH-many}
\end{equation}
We suppose that $f$ is defined by (\ref{f}). For moderate values
of $t$  a solution on the equation (\ref{ord-n-KH-many})
  exists with the asymptotics written in terms of the Hankel function
\begin{equation}
u(s) = s^{-(n-1)/2}H_{(n-1)/2}^{(1)}( m s)   ( 1 + O ( \exp(-2a) ) ), \quad a \to \infty
 \label{estimate-many}
\end{equation}
which is valid outside the moving area $ {( x_{1} - v_{gr} t)^2 /
\Delta_{\parallel}^2 }
    + { (x_{2}^2 + \ldots + x_n^2) / {\Delta}_{\perp}^2 } \ge (a - bm^2)
\label{vicinity-many} $ where $\Delta_{\parallel}$,
${\Delta}_{\perp}$, a are defined above. Localization of the solution
for moderate times follows from the asymptotics of the Hankel
function.

The research is supported by the grant RFBR 0001-00485.

\end{document}